\documentclass[twocolumn, superscriptaddress,nofootinbib,showkeys, aps, prd, longbibliography]{revtex4-1}
\usepackage{amsfonts}           % AMS packages
\usepackage{amssymb}            % ^
\usepackage{amsmath}            % ^
\usepackage{comment}
\usepackage{latexsym}           % For the "normal" symbol (and other funky binary operators)
\usepackage{dcolumn}            % Align table columns on decimal point
\usepackage[dvips]{graphicx}   % For production with eps / jpg figures
\usepackage{enumerate}
\usepackage{epstopdf}           % Can incorporate eps in the above pdftex
\usepackage{fancyhdr}           % Does exactly what it says it will
\usepackage{hyperref}           % Allow for hyperlinks
\hypersetup{
    colorlinks=true,
    linkcolor=blue,
    filecolor=magenta,      
    urlcolor=cyan,
}
\usepackage{setspace}           % Allows us to change spacings
\usepackage{xspace}             % To include spaces after newly defined commands when they are expecting arguments
\usepackage{subfig}
\usepackage{bm}                 % A better "bold math" package -- any math symbol can be emboldened with the command \bm{*}
\usepackage[ulem=normalem]{changes}
\usepackage{url}
\usepackage{color}
\usepackage{lineno}
\usepackage[toc,page]{appendix}
\newcommand{\szz}{\sigma_{\rm zz}}

\newcommand{\rmm}{{\rm m}}
\newcommand{\Lg}{\Lambda}

\newcommand{\rw}{\rightarrow}
\newcommand{\bea}{\begin{eqnarray}}
\newcommand{\eea}{\end{eqnarray}}

\newcommand{\cf}{{\it cf.}~}

\newcommand{\ie}{{\it ie.\,}}
\newcommand{\bfx}{{\bf x}}
\newcommand{\bfz}{{\bf z}}
\newcommand{\bfn}{{\bf n}}
\newcommand{\bfnt}{{\bf \tilde n}}

\newcommand{\bfr}{{\bf r}}

\newcommand{\bft}{{\bf t}}
\newcommand{\bfb}{{\bf b}}

\newcommand{\bfS}{{\bf S}}

\newcommand{\eps}{\epsilon}

\newcommand{\eg}{{\it e.g. }}
\newcommand{\pd}{\partial}
\newcommand{\been}{\begin{enumerate}}
\newcommand{\een}{\end{enumerate}}
\newcommand{\beit}{\begin{itemize}}
\newcommand{\eit}{\end{itemize}}
\newcommand{\bfv}{{\bf v}}
\newcommand{\vs}{{\it vs.~}}

\newcommand{\bP}{\beta^{P}}
\newcommand{\bE}{\beta^{E}}
\newcommand{\curl}{\nabla\times}
\newcommand{\bfFem}{{\bf F^{e-1}}}
\newcommand{\bfFemt}{{\bf F^{e-\tau}}}
\newcommand{\bfT}{{\bf T}}

\newcommand{\cb}{{\bf b}}
\newcommand{\mum}{\mu {\rm m}}
\newcommand{\nm}{{\rm nm}}
\newcommand{\Jij}{J_{\rm ij}}

\newcommand{\dL}[0]{\, \text{d}\ell}
\newcommand{\dV}[0]{\, \text{d}V}
\newcommand{\dS}[0]{\, \text{d}S}

\newcolumntype{L}[1]{>{\raggedright\let\newline\\\arraybackslash\hspace{0pt}}m{#1}}
\newcolumntype{C}[1]{>{\centering\let\newline\\\arraybackslash\hspace{0pt}}m{#1}}
\newcolumntype{R}[1]{>{\raggedleft\let\newline\\\arraybackslash\hspace{0pt}}m{#1}}

\begin{document}

%\linenumbers

%Title.  Two "affiliation" class options are groupedaddress (default) and superscriptaddress
%\title{A Topological Invariant for  Dislocation Networks in Crystals}
\title{The $\Lg$-invariant and topological pathways to influence sub-micron  strength and  crystal plasticity}

\date{\today}

\author{Stefanos Papanikolaou$^\dagger$}
\affiliation{Department of Mechanical Engineering, The West Virginia University}
\affiliation{Department of Physics, The West Virginia University}
\author{Giacomo Po}%$^\dagger$}
\affiliation{Department of Mechanical Engineering, University of Miami}

\let\thefootnote\relax\footnotetext{$^\dagger$stefanos.papanikolaou@mail.wvu.edu}

\begin{abstract}
{%\crd
In small volumes, sample dimensions are known to strongly influence mechanical behavior, especially strength and crystal plasticity. This correlation fades away at the so-called {\it mesoscale}, loosely defined at several micrometers in both experiments and simulations. However, this picture depends on the {\it entanglement} of the initial defect configuration.
In this paper, we study the effect of sample dimensions with a full control on dislocation topology, through the use of a novel observable for dislocation ensembles, the $\Lg$-invariant, that depends only on mutual dislocation linking: It is built on the natural vortex character of dislocations and it has a continuum/discrete correspondence that may assist multiscale modeling descriptions. We investigate arbitrarily complex initial dislocation microstructures in sub-micron-sized pillars, using three-dimensional discrete dislocation dynamics simulations for finite volumes. We demonstrate how to engineer nanoscale dislocation ensembles that appear virtually independent from sample dimensions, either by biased-random dislocation loop deposition or by sequential mechanical loads of compression and torsion.
}
\end{abstract}

\maketitle
% \tableofcontents

{%\crd
%intro
Among the most remarkable aspects of forming processes in metals is the ability to manipulate material strength by ``cold working"~\cite{Asaro:2006fr}. At the heart of this versatile feature lies the ability of crystal defects, especially dislocations~\cite{Anderson:2017aa,Kubin:2013pr}, to  interact collectively, develop entangled microstructures and multiply. Dislocation entanglement has been notoriously believed to control a plethora of phenomena in metallurgy, including work and kinematic hardening, as well as fatigue~\cite{Nabarro:1964ff}. However, the paramount importance of dislocation entanglement only became clear in the study of small finite volumes, by noticing the dramatic effects of its absence~\cite{El-Awady:2009nx,lee2013emergence,Ryu:2015aa,papanikolaou2017obstacles,papanikolaou2017avalanches,Papanikolaou2012}: Crystalline strength drastically increases  when at least one dimension decreases below the so-called {\it mesoscale}, which loosely refers to a few micrometers~\cite{Croes:2011tw,devincre2008dislocation,Bulatov:2006ye,Kocks:2003fu,Mecking:1981lh} where dislocations, conceived as point particles, define their ``mean-free path"~\cite{devincre2008dislocation}. Nevertheless, crystal dislocations are more often than not, loops that may easily extend to the volume boundaries, and thus mutual dislocation topologies may be critical. In addition, far-from-equilibrium basics suggest that mechanical yielding is a {\it transient behavior} that ought to strongly depend on initial conditions, especially in small volumes~\cite{goldenfeld1992lectures}. In this work, we show that dislocation entanglement primarily depends on the dislocation ensemble topology and not a particular lengthscale. We investigate the possible effects of initial conditions by constructing a topological observable of dislocation networks, the $\Lg$-invariant, that is only dependent on mutual loop entanglement. We show that the $\Lg$-invariant can be used to generate arbitrarily complex microstructures, either by deposition or cold-working, rendering a sub-micron volume capable of yielding like a bulk sample. In this way, we may identify topological pathways to manage strength and crystal plasticity and unify critical aspects of multiscale materials modeling.

\begin{figure*}[t]
\includegraphics[angle=0, width=0.9\textwidth]{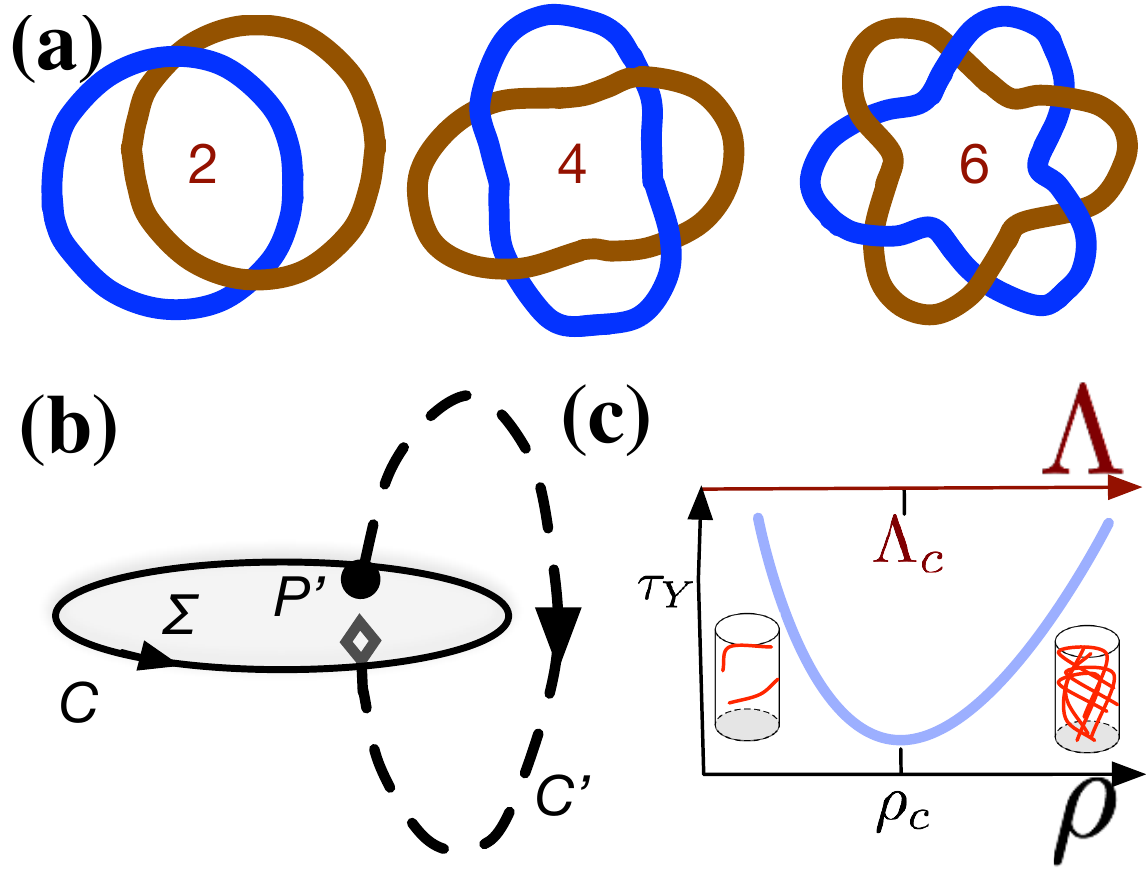}
\caption{{\bf Dislocation Loops and Topology}.
(a) Two dislocation loops on different slip systems can form a topological {\it link} with linking number 2, 4 or 6 depending on the activation of additional latent hardening mechanisms such as double cross-slip.
(b) The {\it linking number} can be directly calculated through a dislocation line double integral, the Gauss integral.
(c) The calculation of a dislocation ensemble linking number (assuming closed dislocation loops) can be performed either in simulations or experiments (Figure).
}
\label{fig:hel1}
\end{figure*}

While  interactions and mechanisms of individual dislocations are fairly understood, crystal plasticity modeling still remains an enormous challenge. For example, the necessary ``back-stress" for theories of kinematic and work hardening~\cite{Asaro:2006fr}, does not yet have a precise microscopic definition~\cite{El-Azab:00,chen2010bending}.  In fact, the great complexity of crystal plasticity theories originates in that dislocation ensembles are more akin to a ``bird's nest", rather than a set of separate small and simpler elementary bodies~\cite{Cottrell:2002cj}. In sub-micron-sized volumes, where it becomes tough to fit such a nest, common plasticity practically disappears, giving its place to uncommon size effects and stochasticity in the mechanical response~\cite{dimiduk2006scale,uchic2009plasticity,papanikolaou2017avalanches}. In association, cold-working a micropillar before testing~\cite{El-Awady:2011aa} may turn size effects into Taylor hardening (\cf Fig.~1(c)), as dislocation density increases~\cite{el2015unravelling}. However, another plausible interpretation is that dislocation {\it complexity} is the key, in a small finite volume, that may unlock common plasticity. In such a scenario, the inflection point $\Lg_c$ has a fundamental importance in terms of topological complexity, possibly revealing how many dislocation ``twigs" need to intertwine to start behaving as a bird's nest.

\begin{figure*}[tbh]
\includegraphics[angle=0, width=0.9\textwidth]{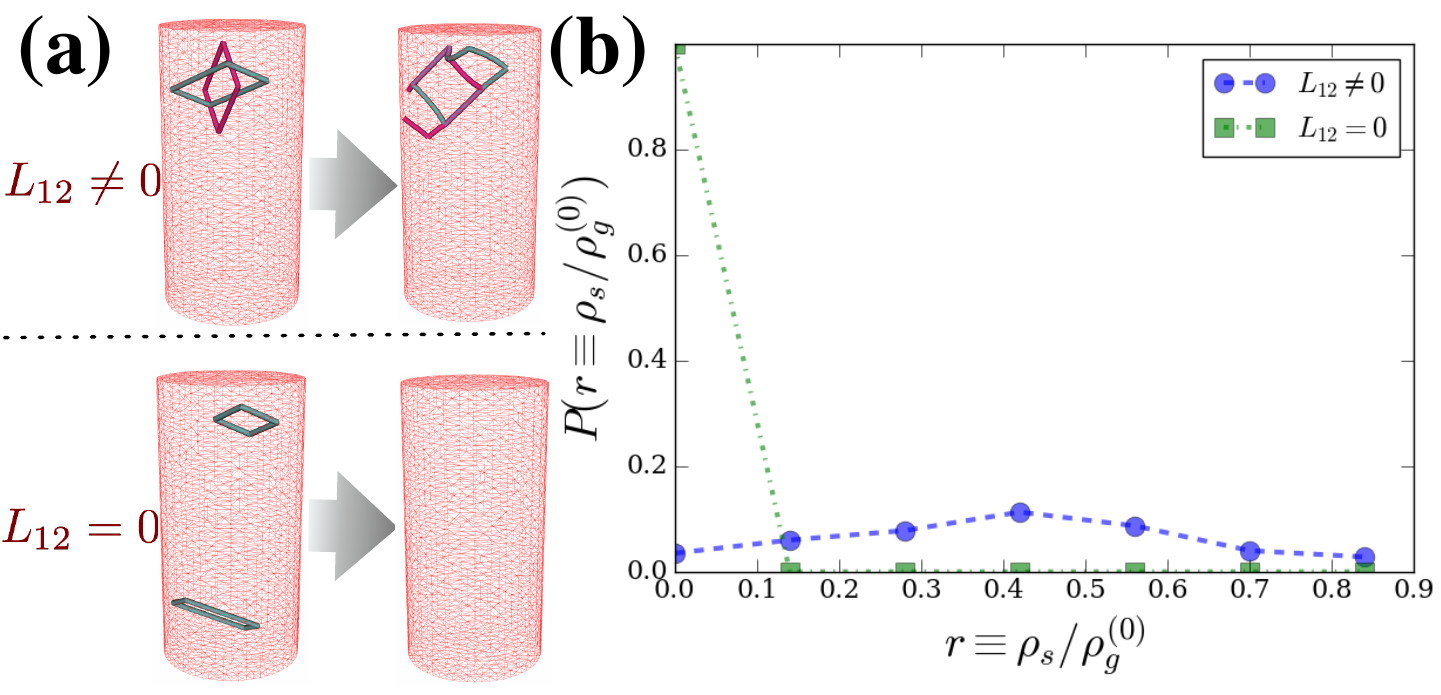}
\caption{{\bf Energetics \& Statistics}: (a) Initial conditions of i) two prismatic dislocation loops 1 and 2, with randomly selected Burgers vectors but finite $L_{12}$, ii)  two prismatic dislocation loops 1 and 2, with randomly selected Burgers vectors but zero $L_{12}$. (b) Probability of forming sessile dislocation junctions through 2 randomly placed prismatic loops with or without linking.}
\label{fig:hel2}
\end{figure*}

We present a novel approach to characterize and engineer dislocation entanglement  that naturally translates into continuum and large-deformation descriptions of dislocation ensembles. Our basis is the construction of a scalar volume observable, dubbed $\Lg$-invariant, which may be shown to have special topological properties, thus leaping beyond the distortion~(elastic/plastic) or dislocation density tensors~\cite{Hirth_1}. The purpose of $\Lg$ is to count and sum the linking number of each pair of dislocation loops across the ensemble. 

The  {\it Linking Number} for two dislocation loops $i$ and $j$ is defined as $L_{ij}=\frac{1}{4\pi}\int_{C_1}\int_{C_2} d\Omega (\bfr_1,\bfr_2)\equiv\frac{1}{4\pi}\int_{C_1}\int_{C_2}\frac{(d\bfr_2\times\bfr_1)\bfr_{12}}{r_{12}^3}$~\cite{Murasugi:2007kt,Kauffman:1991kl} and it is a plausible way to define the mutual entanglement of a pair of dislocation loops (\cf Fig.~\ref{fig:hel1}(b)). A linking number of 2 is typical for crossing dislocations in different slip systems, while higher linking numbers require additional consecutive mechanisms such as consercutive double cross-slip events (\cf Fig.~\ref{fig:hel1}(a)). The topological character of the linking number originates in that it does not depend on local line distortions, thus it does not explicitly relate to dislocation {\it length} density. In this way, it complements common dislocation network observables.

The $\Lg$-invariant in a crystal of Burgers vector magnitude $b$ is defined as,
\bea
\Lg = \frac{1}{b^2}\int d^3x \bE \cdot (\curl\bE)=\frac{1}{b^2}\int d^3x \bE_{ij} \alpha_{ij}
\eea
where $\alpha$ is the Nye dislocation density tensor and the elastic distortion $\bE$ combines with the plastic distortion $\bP$ to give $\bP+\bE=grad(u)$, where $u$ is the displacement field due to deformation \cite{Asaro:2006fr,Anderson:2017aa}, ultimately satisfying on a closed crystal boundary $\Gamma$
\bea
\alpha=\curl\bE =-\curl\bP = \delta_\Gamma\otimes \vec{b}
\eea
This defining vortex character of dislocations signifies that dislocations maintain a {\it loop} character that may not end within the crystal. By using this feature, one may show (see Appendix A)~\cite{Moffatt:2013lq,Moffatt:1969db} that
\bea
\Lg=-\frac{1}{b^2}\sum_{i,j}L_{ij}\cb_i\cdot \cb_j
\eea
where $L_{ij}$ is the linking number between loops $i$ and $j$, while $\cb_i$  and $\cb_j$ are their respective Burgers vectors. 

\begin{figure*}[t]
\includegraphics[angle=0, width=0.9\textwidth]{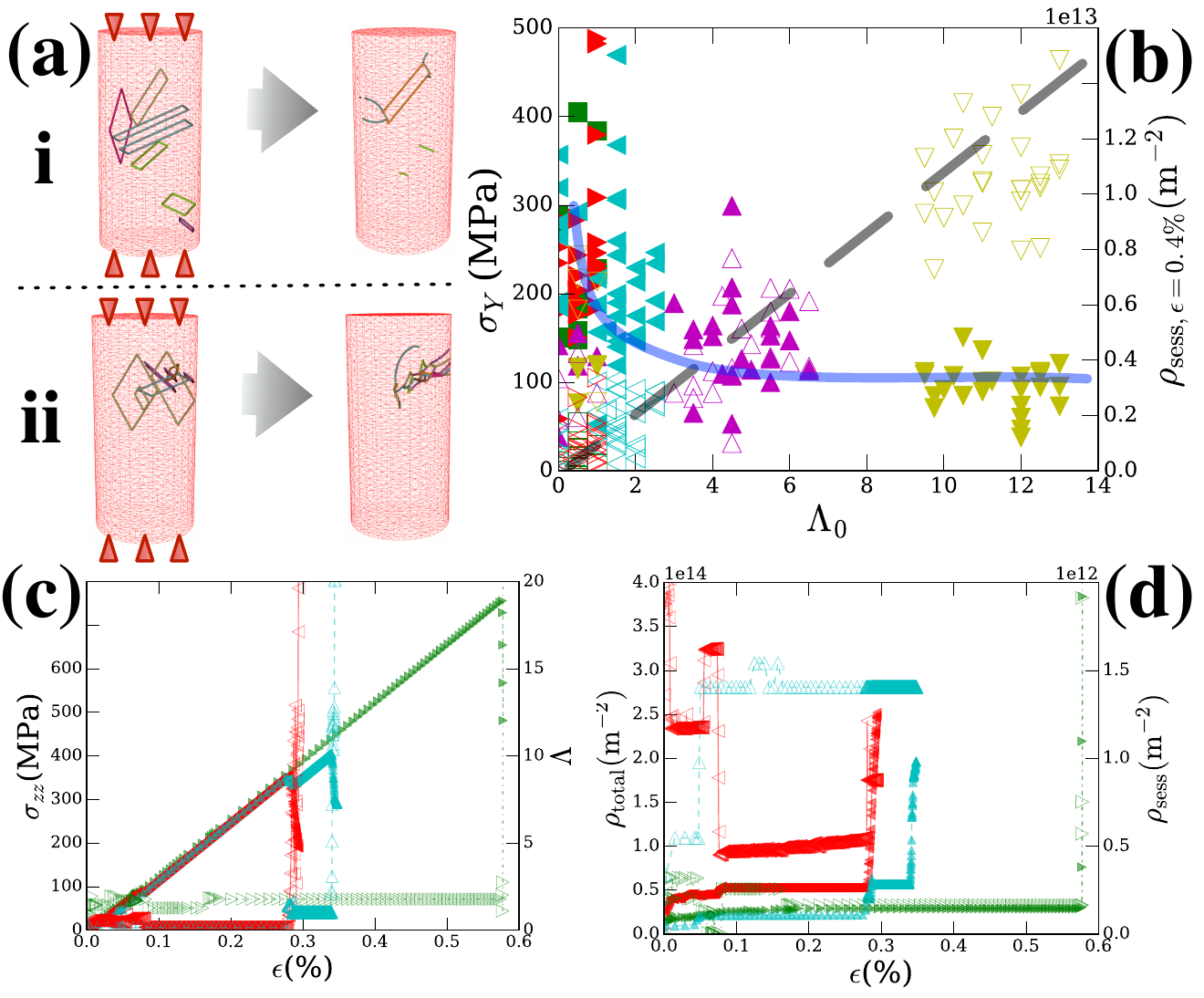}
\caption{{\bf Depositing\&Testing Entanglement -- Uniaxial Compression in Designed Dislocation Configurations $\&$ Size In-dependence}
(a) Random deposition and uniaxial compression of prismatic dislocation loops in a finite mesh of $R=0.3\mum$ in (i) an unbiased or (ii) $\Lg$-biased manner. (shown compression to $0.01\%$ strain for $\rho_0=2\times 10^{13}/\rmm^2$)
(b) Nanopillar yield stress (solid symbols) and sessile dislocation density (open symbols) at $0.2\%$ strain as function of initial $\Lg_0$.  Every point corresponds to a uniaxial compression simulation: ($\rho_0:\{\blacksquare:10^{13}/\rmm^2\}$, $\{\blacktriangleright:2\times10^{13}/\rmm^2\}$, $\{\blacktriangleleft:4\times10^{13}/\rmm^2\}$, $\{\blacktriangle:8\times10^{13}/\rmm^2\}$, $\{\blacktriangledown:10^{14}/\rmm^2\}$). Also, three sample cases are shown for (c) loading stress $\szz$ and $\Lg$ \vs $\eps$, and (d) total and sessile dislocation density \vs $\eps$. Sessile dislocation density increases roughly linearly with $\Lg_0$ while yield stress quickly saturates. ($\{\blacktriangle:8\times10^{12}/\rmm^2\}$, $\{\blacktriangleright:10^{13}/\rmm^2\}$, $\{\blacktriangleleft:2\times10^{13}/\rmm^2\}$)
}
\label{fig:hel3}
\end{figure*}

The primary usefulness of $\Lg$ is its capacity of predicting the onset of dislocation multiplication through dislocation junction formation and associated mechanisms. The connection between finite $\Lg$ and junction formation can be seen in two ways: First, the absence of linking ($L_{ij}=0$) leads to a negligible {\it statistical} probability for junction formation, especially in small finite volumes (\cf Fig.~\ref{fig:hel2}). Second, the dependence of $\Lg$ on the linking loops' Burgers vectors' dot product point directly to a junction-formation energetic connection to the Frank rule. 
A way to realize different possibilities is to consider prismatic dislocation loops, which have been connected to hardening effects in various circumstances~\cite{Puschl:2002wv,Saada:1962yt}. If one considers two prismatic dislocation loops randomly placed in a small finite volume, then the histogram of {\it sessile} dislocation junction formation displays significant junction length formation
for both signs of $\Lg$, with a wide probability distribution (see Fig.~\ref{fig:hel2}) that remarkably overwhelms the unlinking case. It is worth noting that it has been common to consider segment-based junction formation arguments~\cite{Wickham:1999qd, Madec:2002rc} (\ie nearby straight lines), but the scenario of two dislocation loops that intertwine is significantly different since all possible signs of dislocation interactions are present in such a linked dislocation pair. Thus, it is natural to expect a significantly larger {\it statistical} preference towards sessile junction formation for linked loops than two nearby straight lines~\cite{Shenoy:2000xw,Madec:2002kn}

\begin{figure*}[tbh]
\includegraphics[width=0.95\textwidth]{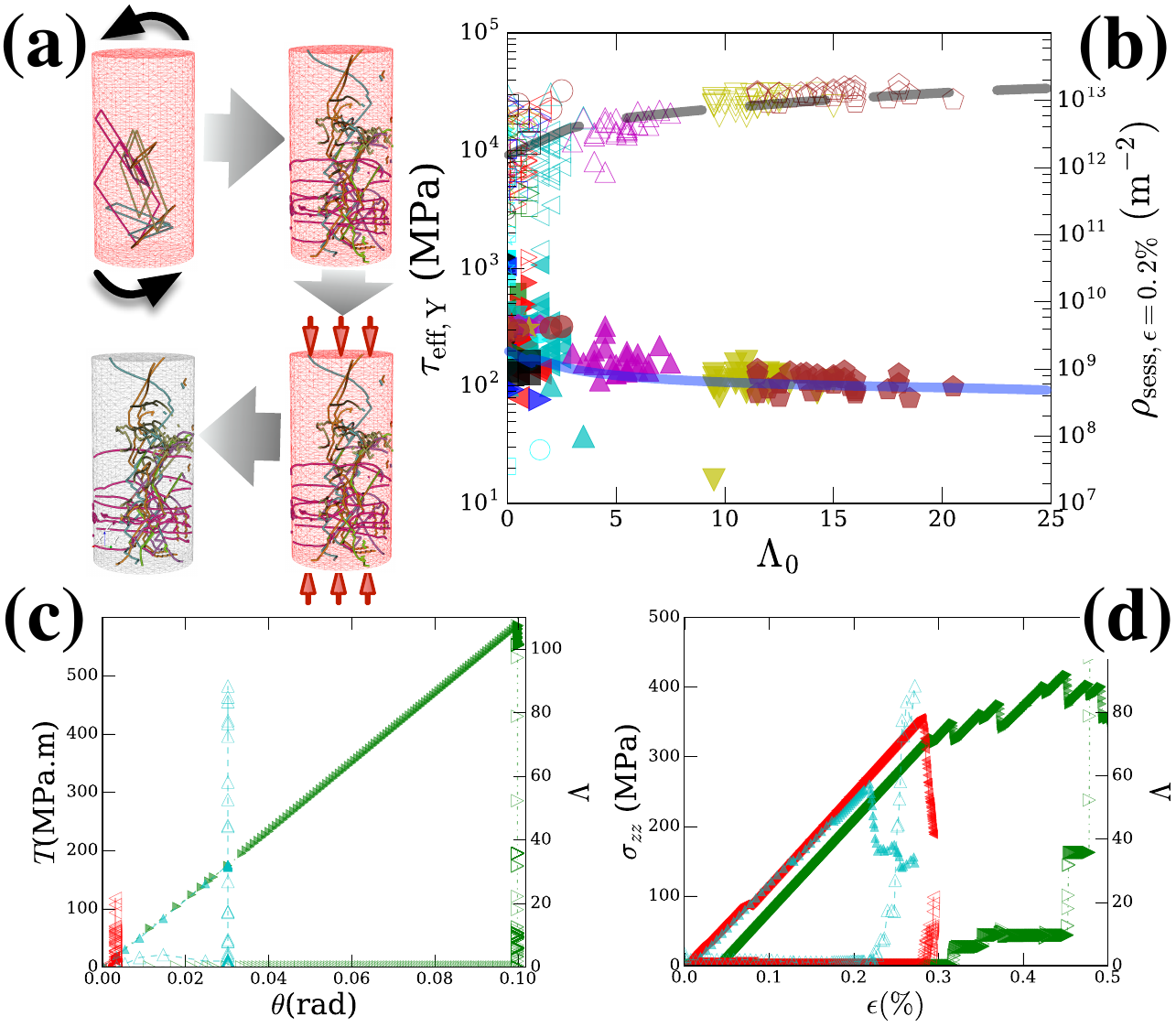}
\caption{{\bf Engineering Entanglement by Cold Working: Pure Torsion Followed by Uniaxial Compression}. 
(a) A protocol is followed where a dislocation loop configuration at density $\rho_0$ is prepared,  torsion is applied on top/bottom pillar surfaces up to pre-chosen torsion angle $\theta_0$, and then uniaxial compression is applied towards yield. The configurations at each step of the process are shown for a particular configuration with $\rho_0=3\times10^{13}/\rmm^2$.
(b) Yield stress \vs $\Lg_0$ for a variety of initial conditions and torsions $\rho_0\in(5\times10^{12}, 2\times10^{14})$ and $T\in\{10^{-5},0.1\}$. Three characteristic cases are shown for
(c) Torsion $T$ (filled symbols) and $\Lg$ (open symbols) \vs torsion angle $\theta$ and (d) subsequent loading stress $\szz$ (filled symbols) and $\Lg$ (open symbols) \vs axial strain $\eps$. $\{\rho_0,\theta_0\}\rw$\Big($\blacktriangleright:[10^{13}/\rmm^2,0.1]$, $\blacktriangleleft:[2\times10^{13}/\rmm^2,3\times10^{-3}]$, $\blacktriangle:[2\times10^{13}/\rmm^2,3\times10^{-2}]\Big)$.
}
\label{fig:hel4}
\end{figure*}

The statistical finding for the fate of two linked loops has significant consequences for the behavior of collective dislocation networks. By using a numerical algorithm that explicitly tracks dislocation loops~\cite{modelib} and their mutual linking numbers throughout the dislocation dynamics simulation~\cite{Ghoniem:88a,Arsenlis:2007cj,vandergiessen1995,WeygandNeedleman01,po2014recent,modelib}, we are able to arbitrarily tune the complexity of the initially deposited dislocation configuration. We consider the geometry of a cylindrical nanopillar finite element mesh of diameter $D=2000b\sim600$nm for single crystalline Cu FCC with $\bfb=0.185\nm$. We generate a wealth of prismatic loop initial conditions by depositing prismatic dislocation loops of randomly selected Burgers vectors in the nanopillar until a target dislocation density $\rho_0$ is reached (see \eg Fig.~\ref{fig:hel3}(a) for $\rho_0=10^{13}/{\rm m}^2$). Furthermore, by biasing the deposition of prismatic loops towards collectively increasing $\Lg$ magnitude, we acquire complete control on the investigation of dislocation topological effects. 

The effect of topologically rich initial conditions is drastic in causing statistical {\it size independence} in material properties such as the compressive yield strength (\cf Fig.~\ref{fig:hel3}(b)). For large initial $\Lg_0$ (in this work, we focus on $\Lg$'s magnitude), in fact, the strength of the pillars is dominated by the microstructure entanglement as opposed to the sample size, leading to a linear increase of the sessile dislocation density at an arbitrarily chosen $0.4\%$ finite strain $(\rho_{sess}\sim\Lg_0)$. While this is only evidence of a scaling relation between $\Lg_0$ and dislocation densities in small finite volumes, we expect that the scaling relationship $\rho_{sess}(\eps)\sim \Lg_{0}^{\delta}$ for $\delta\geq0$ generically holds for any dislocation ensemble.  Plastic yielding in these systems is accompanied by large $\Lg$-``avalanches" (\cf Fig.~\ref{fig:hel3}(c)), caused by large increase in the density of {\it sessile} (\cf Fig.~\ref{fig:hel3}(d)), and subsequent multiplication to increase the total dislocation density (\cf Fig.~\ref{fig:hel3}(d)). Overall, this behavior should be contrasted to the typically observed size-dependent one in analogously small finite volumes~\cite{Uchic2009}, which in our simulations can be seen only for $\Lg_0<1.8$.

%We demonstrate this effect by considering Discrete Dislocation Dynamics (DDD) simulations~ in $2000\bfb$-diameter and $4000\bfb$-height pillars, where the Burgers vector $\bfb=0.185\nm$, and finite volume boundary conditions are implemented~. It is worth noting that our DDD algorithm maintains and tracks the dislocation loop structures {\it throughout} the simulation. Our samples are generated for a Cu FCC crystal in a finite element mesh. We consider initial conditions where randomly sized and placed prismatic dislocation loops are added to the volume only if the addition increases collectively $\Lg$ (\cf Fig.~\ref{fig:hel3}).
}

Besides artificial deposition of initial dislocation configurations with dramatic effects on mechanical properties, the topological character of $\Lg$ may guide us towards generating large-entanglement structures through particularly {\it efficient} mechanical loading paths. These loading paths can be also predicted through modeling of latent hardening for particular crystalline structure~\cite{Lagneborg:1973vh}. The key towards identifying such loading paths is the calculation of the $\Lg$-dynamics. For a large class of continuum dislocation theories that satisfy global Burgers vector conservation and Orowan's law of collective dislocation motion, it may be shown that (Appendix B)
\bea
\frac{\pd \Lg}{\pd t} = \frac{1}{b^2}
\int_{\pd V}dS_j  \eps_{jkm} \bE_{kl} J_{ml}
\label{eq:loadm}
\eea
where the typical assumption of overdamped dynamics is considered and $J_{ij}=\eps_{ikm}F_m\alpha_{kj}$, where $F_m$ is the PK force on the dislocation density $\alpha_{kj}$. The tensorial character of the right-hand side of Eq.~\ref{eq:loadm} implies that only particular loading directions can increase $\Lg$, a fact well known from studies of latent hardening in crystal plasticity~\cite{BDevincreKubin94,devincre2008dislocation,Kocks66,Takeuchi:1975ak}.  A  physically intuitive example is the uniaxial compression of pre-torsioned specimens, where dislocation flow $\Jij$ may be assumed in a pre-strained environment to a torsion angle $\theta_0$, with induced $\bE$ along the cylindrical $\theta$ direction (\ie $\pd_xu_z$ and $\pd_yu_z$ are non-zero) while $\bft\times\bfv$ for an x-y gliding dislocation would be along $\bfz$ during compression. This combination of indices gives a concrete contribution to the right-hand-side integral of Eq.~\ref{eq:loadm}. Physically, in the idealized continuum cases, torsion induces geometrically imposed screw dislocations along the torsion axis. It is natural to expect that torsion-induced screw dislocations along the loading axis would tangle with horizontal-moving slip during subsequent compression, and this is precisely what Eq.~\ref{eq:loadm} is predicting.

To confirm the approach towards the generation of $\Lg$, we perform explicit 3D-DDD sequential-loading simulations of submicron-sized pillars with various initial dislocation densities $\rho_0$. We vary $\rho_0$ from $5\times 10^{12}$ to $3\times 10^{14}$. As shown in Fig.~\ref{fig:hel4}(a), the application of a finite amount of torsion (which may not be necessarily large enough to induce plasticity) on the top/bottom surfaces up to a target torsion angle $\theta_0$ ($\in (0,0.1)$rad), leads to a highly extended dislocation configuration that generates large entanglement when it is followed by uniaxial compression, as it is witnessed by the increase of $\Lg$. Characteristically, if one calculates the combined-loading effective stress, then the yield stress is remarkably size-independent (\cf Fig.~\ref{fig:hel4}(b)) and the saturation level for the sessile dislocation density is easily at virtually any case, even with miniscule initial $\Lg$ and initial dislocation density. As one may see in particular example cases (\cf Fig.~\ref{fig:hel4}(c),(d)), the application of torsion is followed with a dramatic increase of $\Lg$, as suspected by our developed topological intuition.

The usefulness of $\Lg$ is not limited to the characterization and prediction of discrete and entangled dislocation networks, but also it extends to represent a unique {\it discrete-continuum} link that can be directly calculated in both the discrete and the continuum worlds. In the continuum, it is just needed to properly estimate a $\bE$-dependent volume integral. Its topological origin also allows us to write the correspondent of $\Lg$ for large deformations. Following Ref.~\cite{Cermelli:2001pz} a large deformation generalization may be shown to be, (Appendix C)
\bea
\Lg=\frac{1}{b^2}\int J^e \bfFem \cdot (\curl\bfFem)^{-\tau} \cdot \bfFemt d\tilde{V}
\label{eq:large-deform}
\eea
In this way, $\Lg$'s information may be instrumental for extending dislocations' role towards multiscale modeling of large scales and large deformations.

In conclusion, we presented a topological approach to investigate dislocation entanglement~\cite{Devincre:2006xq} and latent hardening in crystals. We find that the manipulation of initially prepared dislocation configurations' topological complexity can generate {\it size independent} crystal plasticity even in nanoscale volumes, that are believed to be {\it intrinsically} size-dependent~\cite{Voyiadjis:2017et,Weinberger:2011qd}. This leap of understanding on the manipulation and control of dislocation networks, may allow for optimization and design of multi-axial, sequential cold-working pathways in metallurgy.\cite{Hansen:1995ca,Hong:2013sy}

\begin{acknowledgments}
We would like to thank J. Bassani, N. Ghoniem, A. Reid, A. Rollett and E. Van der Giessen for inspiring discussions and comments. This work is supported through the award DOE-BES DE-SC0014109 (SP). This work benefited greatly from the facilities and staff of the Super Computing System (Spruce Knob) at West Virginia University.
\end{acknowledgments}

%\bibliography{plasticity}

\begin{appendices}
\section{$\Lg$ and Linking Numbers}
\bea
\oint_{\partial \mathcal{S}} \epsilon_{klm}T dL_k=\int_\mathcal{S}\left(T_{,m}dS_{l}-T_{,l}dS_m\right)
\eea

\bea
\oint_{\partial \mathcal{S}} T dL_i=\int_\mathcal{S}\epsilon_{ijk}T_{,k}dS_{j}
\eea

\bea
\frac{1}{4\pi}\oint\oint\frac{\bm r_1-\bm r_2}{\|\bm r_1-\bm r_2\|^3}\cdot (d\bm r_1\times d\bm r_2)=\nonumber\\
=\left(\frac{1}{R}\right)_{,j}\epsilon_{jkl}dx^{(A)}_kdx^{(B)}_l\nonumber\\
=\frac{1}{2}R_{,ppj}\, \epsilon_{jkl}dx^{(A)}_kdx^{(B)}_l
\eea

\bea
b^2\Lg&=&\int_{\mathbb{R}^3}\beta^E_{ij}\alpha_{ij}\dV\nonumber\\
&=&\int_{\mathbb{R}^3}\beta^E_{ij}\sum_A\oint_{\mathcal{L}^A}\delta(\bm x-\bm x')b_i^A\dL_j'\dV\nonumber\\
&=&\sum_Ab_i^A\oint_{\mathcal{L}^A}\beta^E_{ij} \dL_j\nonumber\\
&=&\sum_Ab_i^A\oint_{\mathcal{L}^A}u_{i,j} \dL_j\mp\sum_Ab_i^A\oint_{\mathcal{L}^A}\beta^P_{ij} \dL_j\nonumber\\
&=&\sum_Ab_i^A\oint_{\mathcal{L}^A}\sum_B\int_{\mathcal{S}^B}b^B_i \delta(\bm x-\bm x')\dS_j'\dL_j\nonumber\\
&=&-\frac{1}{4\pi}\sum_A\sum_Bb_i^Ab^B_i\oint_{\mathcal{L}^A}\int_{\mathcal{S}^B} \left(\frac{1}{R}\right)_{,pp}\dS_j'\dL_j\nonumber\\
&=&-\frac{1}{4\pi}\sum_A\sum_Bb_i^Ab^B_i\oint_{\mathcal{L}^A}\int_{\mathcal{S}^B} \left(\frac{1}{R}\right)_{,pj}\dS_p'\dL_j\nonumber\\
&&-\frac{1}{4\pi}\sum_A\sum_Bb_i^Ab^B_i\oint_{\mathcal{L}^A}\oint_{\mathcal{L}^B} \epsilon_{kjp}\left(\frac{1}{R}\right)_{,p}\dL_k'\dL_j\nonumber\\
&=&-\frac{1}{4\pi}\sum_A\sum_Bb_i^Ab^B_i\oint_{\mathcal{L}^A}\oint_{\mathcal{L}^B} \epsilon_{kjp}\left(\frac{1}{R}\right)_{,p}\dL_k'\dL_j' \nonumber\\
&=&-\sum_{A,B}L_{AB}\bfb_A \cdot \bfb_B
\eea

\section{The $\Lg$ Dynamics}
{ For x-y torsion $\beta_E \sim \hat\theta$ (primarily) in cylindrical coordinates, since most dislocations (screw) are lying along z-axis. Moreover, dynamics during compression in a torsioned specimen implies that $J\sim -\hat r$ in cylindrical coordinates. (to check it again)
}

In order to be careful, one needs to carefully perform these calculations using the proper tensorial indices. In this context, it is important to clarify that:
\been
\item For $\alpha_{ij}$: $i$ indexes the dislocation line vector $\bft$ that the dislocation is tangent to, while $j$ indexes the Burgers vector $\bfb$.
\item For $J_{ij}$ and a single dislocation line moving with velocity $\bfv$, $J_{ij}=\eps_{ikm}t_kb_jv_m\delta(\bfx)$ at some location $\bfx$. So, $i$ labels the cross product $\bft\times \bfv$.
\item Also, for $J_{ij}$ (if overdamped dynamics is considered): it is written as $J_{ij}=\eps_{ikm}F_m\alpha_{kj}$, where $F_m$ is the PK force on the dislocation density $\alpha_{kj}$ ($F\times\alpha$ making sure that a dislocation density moves perpendicularly to the dislocation line vector).
\item For $\beta^{E}_{ij}$: $i$ labels the strain direction, while $j$ the component of the displacement vector.
\item We need to assume a particular dynamics law for the elastic and plastic distortion in order to proceed. Assuming the Nye dislocation tensor,
\bea
\alpha_{ij} = -\eps_{ilm}\pd_l \bP_{mj} = \eps_{ilm}\pd_l \bE_{mj}
\eea
we can assume a generic conservation law for the Burgers' vector:
\bea
\pd_t\alpha_{ij} = -\eps_{ilm}\pd_l J_{mj}
\label{eq:balphaeq}
\eea 
which also implies,
\bea
\pd_t\bE_{ij} = - J_{ij}
\label{eq:bEeq}
\eea
\een

The ultimate target is to understand and predict which loading modes can lead to a large increase of the dislocation helicity, and consequently the elastic energy of the crystal. Assuming that the volume V is fixed, with $\pd V$ a boundary surface, then, if the helicity is defined as:
\bea
\Lg=\frac{1}{b^2}\int_VdV\bE_{kl}\eps_{kmn}\pd_m\bE_{nl}
\eea
or
\bea
\Lg=\frac{1}{b^2}\int_VdV\bE_{kl}\alpha_{kl}
\eea

then the temporal variation of dislocation helicity is derived by direct differentiation:
\bea
b^2\frac{\pd \Lg}{\pd t} = \int_VdV \frac{\pd \bE_{kl}}{\pd t} \alpha_{kl} + \int_VdV \bE_{kl} \frac{\pd \alpha_{kl}}{\pd t}
\eea
Now we can use Eqs.~\ref{eq:balphaeq} and \ref{eq:bEeq}, showing that
\bea
b^2\frac{\pd \Lg}{\pd t} = 
-\int_VdV J_{kl} \alpha_{kl} - 
\int_VdV \bE_{kl} \eps_{kjm}\pd_j J_{ml} 
\eea
Now, we may use integration by parts in the second integral:
\bea
b^2\frac{\pd \Lg}{\pd t} = 
-\int_VdV J_{kl} \alpha_{kl} - 
\int_{\pd V}dS_j \bE_{kl} \eps_{kjm} J_{ml} +\nonumber\\
\int_{ V}dV \pd_j\bE_{kl} \eps_{kjm} J_{ml}
\eea

which is equal to:

\bea
b^2\frac{\pd \Lg}{\pd t} = 
-\int_VdV J_{kl} \alpha_{kl} - 
\int_{ V}dV \alpha_{ml} J_{ml} -\nonumber\\
\int_{\pd V}dS_j \bE_{kl} \eps_{kjm} J_{ml}
\eea

or

\bea
b^2\frac{\pd \Lg}{\pd t} = 
-2\int_VdV J_{kl} \alpha_{kl} +
\int_{\pd V}dS_j  \eps_{jkm} \bE_{kl} J_{ml}
\label{eq:load}
\eea
The first integral (volume) is identically zero for the conservative dynamics, since $J_{kl}\alpha_{kl}=\eps_{kij}\alpha_{il}F_j\alpha_{kl}\equiv \eps_{kij}\alpha_{il}\alpha_{kl}F_j$ (zero by reversing k, i). {Thus, it is proven that dislocation helicity is conserved by volume dynamics}.

The second integral is equivalent to:
\bea
\int_{\pd V} d\bfS\cdot (\bE\times J)
\label{eq:hel-evol}
\eea

\section{$\Lg$ at large deformations}

In the deformed configuration $\bfx$ (also labeled by  above symbols), one can still define simply the burgers vector as a smooth loop/surface integral:
\bea
\bfb = \int_{\pd S}\bfFem d\bfx = \int_{S}(\curl \bfFem)^\tau\bfnt d\tilde A
\eea
with $\bfFem=({\bf F^e})^{-1}$ and where $\bfn$ is the vector normal to the area S. Also, it is important to remember that,
\bea
(\curl \bfT)_{ij} = \eps_{irs} \frac{\pd T_{js}}{\pd x_{r}}
\eea
with respect to the orthonormal basis vectors.

At large deformations, it is important to remember that the deformed coordinates are connected to the {\it reference} configuration through the relation:
\bea
\tilde\bfn d\tilde A= J^e \bfFemt\bfn dA 
\eea

Regarding the $\Lg-$invariant, it is transparent how to extend our definitions in the large deformation regime, in the deformed coordinates. Namely, one can define,
\bea
\Lg=\frac{1}{b^2}\int  \bfFem \cdot (\curl\bfFem)^{-\tau} d\tilde{V}
\eea
and in the {\bf reference} frame:
\bea
\Lg=\frac{1}{b^2}\int J^e \bfFem \cdot (\curl\bfFem)^{-\tau} \cdot \bfFemt d\tilde{V}
\label{eq:large-deform}
\eea
Eq.~\ref{eq:large-deform} represents the large deformation definition of the $\Lg-$invariant.

\end{appendices}

\end{document}